\documentclass[conference]{IEEEtran}
\IEEEoverridecommandlockouts
\usepackage{cite}
\usepackage{amsmath,amssymb,amsfonts}
\usepackage{algorithmic}
\usepackage{graphicx}
\usepackage{textcomp}
\usepackage{xcolor}
\usepackage{float}
\usepackage{subfig}
\usepackage{adjustbox}
\usepackage{subfloat}
\usepackage{siunitx}
\usepackage{booktabs}
\usepackage{footmisc}
\usepackage{url}
\usepackage{balance}

\def\BibTeX{{\rm B\kern-.05em{\sc i\kern-.025em b}\kern-.08em
    T\kern-.1667em\lower.7ex\hbox{E}\kern-.125emX}}
\usepackage{balance}  
\usepackage[algo2e]{algorithm2e}
\def\algbackskip{\hskip-\ALG@thistlm}
\usepackage{algorithm}

\begin{document}

\title{Truth Discovery in Sequence Labels from Crowds\\
}

\author{\IEEEauthorblockN{Nasim Sabetpour $^*$}
\IEEEauthorblockA{
\textit{Iowa State University}\\
Ames, Iowa, USA \\
nasim@iastate.edu}
\and
\and
\IEEEauthorblockN{Adithya Kulkarni  $^*$}
\thanks{$^{*}$ The first two authors contributed equally to this work.}%
\IEEEauthorblockA{
\textit{Iowa State University}\\
Ames, Iowa, USA \\
aditkulk@iastate.edu}
\and
\IEEEauthorblockN{Sihong Xie}
\IEEEauthorblockA{
\textit{Lehigh University}\\
Bethlehem, PA, USA \\
six316@lehigh.edu}
\and
\IEEEauthorblockN{Qi Li}
\IEEEauthorblockA{
\textit{Iowa State University}\\
Ames, Iowa, USA \\
qli@iastate.edu}}
\maketitle
\begin{abstract}
Annotation quality and quantity positively affect the learning performance of sequence labeling, a vital task in Natural Language Processing. Hiring domain experts to annotate a corpus is very costly in terms of money and time. Crowdsourcing platforms, such as Amazon Mechanical Turk (AMT), have been deployed to assist in this purpose. However, the annotations collected this way are prone to human errors due to the lack of expertise of the crowd workers. Existing literature in annotation aggregation assumes that annotations are independent and thus faces challenges when handling the sequential label aggregation tasks with complex dependencies. To conquer the challenges, we propose an optimization-based method that infers the ground truth labels using annotations provided by workers for sequential labeling tasks. The proposed Aggregation method for Sequential Labels from Crowds ($AggSLC$) jointly considers the characteristics of sequential labeling tasks, workers' reliabilities, and advanced machine learning techniques. Theoretical analysis on the algorithm's convergence further demonstrates that the proposed $AggSLC$ halts after a finite number of iterations. We evaluate $AggSLC$ on different crowdsourced datasets for Named Entity Recognition (NER) tasks and Information Extraction tasks in biomedical (PICO), as well as a simulated dataset. Our results show that the proposed method outperforms the state-of-the-art aggregation methods. To achieve insights into the framework, we study the effectiveness of $AggSLC$'s components through ablation studies. 
\end{abstract}

\section{Introduction}
\noindent Online crowdsourcing platforms, such as Amazon Mechanical Turk (AMT), have provided an approachable and low-priced resource to obtain annotated data \cite{2008cheap,novotney2010cheap}. Crowdsourcing platforms distribute annotation tasks among workers to be done in a considerably smaller amount of time. It has changed the process of building the annotated data for various machine learning tasks. Although crowdsourcing platforms are leveraged to help collect annotated data, the labels/annotations provided by workers may be noisier comparing with labels/annotations supplied by experts.

One common practice is to hire multiple workers to annotate the same set of data instances so that errors from individual workers can be detected and corrected. For different types of annotations, different post hoc strategies are applied to ensure the annotation quality. For example, the disagreement in the annotations can be resolved by taking the label with the highest votes as the final result, or a label would be regarded as correct if it receives a specific count of votes. These naive data aggregation methods assume that all workers are equally reliable, which, unfortunately, may not be valid in real-world applications. Advanced aggregation methods are proposed to learn the workers' reliability degrees from their annotations and improve the aggregation results \cite{li2016survey, zheng2017truth}. However, most existing methods assume that data instances are independent, leading to performance loss for tasks such as sequential labeling, a common NLP task, where tokens in a sentence are not independent. Moreover, when the dependencies among data instances are considered, it can be easier to infer correct labels because the dependencies can provide additional information to estimate the correctness of labels. 

Multiple aggregation approaches are recently proposed to handle the particular characteristics of sequential labeling tasks, where tokens in the sentence have complex dependencies \cite{rodrigues2014sequence, simpson-gurevych-2019-bayesian, nguyen2017aggregating}. In these works, probabilistic models are adapted to model the workers' labeling behavior and the dependencies between adjacent tokens. However, there are some drawbacks to the probabilistic models. First, they have solid statistical assumptions when modeling the sequence annotations, limiting the flexibility of the models. Second, these models need to infer complex parameters, making it hard to interpret the relations between workers' quality and token's true labels. Third, these aggregation methods cannot fully unleash the power of state-of-the-art machine learning in sequential labeling tasks. Recently, an optimization approach has been proposed to aggregate sequence labels with complex dependencies \cite{sabetpour2020optsla}; however, the proposed model handles dependencies between tokens by a unidirectional function, limiting the benefits of incorporating consistency function into the framework.

To address these challenges, we propose an optimization framework to improve aggregation performance. For a token, its label is more likely to be correct if provided by reliable workers, agreed by a highly accurate deep learning model, and coherent with the labels around it. Building upon these observations, the proposed method contains three modules: the aggregation, prediction, and consistency modules. The overall goal is to find the true sequence of labels for each sentence to minimize the deviants from all three modules jointly.  

The aggregation module considers workers' labels, reliability, and aggregation confidence to infer the true labels for each token. This module minimizes the weighted loss of the aggregation results with the crowdsourced annotations. The prediction module further applies the state-of-the-art machine learning approach to incorporate corpora features for the sequential labeling task. Unlike existing learning from crowdsourced label techniques that use all noisy labels \cite{rodrigues2018deep,lan2019learning}, $AggSLC$ chooses sentences with high confidence from the aggregation module to ensure high-quality training data. The machine learning model in the prediction module is incrementally trained with the iteratively updated aggregation results. The prediction from the machine learning model, in turn, provides additional annotations to the aggregation module. As a result, the prediction module and the aggregation module can mutually enhance each other. The consistency module models the label dependencies in the sequential labeling task. This module penalizes labels that violate the sequential labeling rules. We propose a bidirectional approach inspired by the Viterbi algorithm \cite{nguyen2007comparisons} to keep the consistency following the sequential labeling rules. A gradient descent approach is adopted as the solution to the optimization problem. The model parameters are updated iteratively. We further link the proposed solution to the EM algorithm and prove the convergence of the proposed solution. Empirically, the convergence can be achieved in several iterations. 

In summary, we present a general framework for the sequential annotation aggregation problem. $AggSLC$ jointly considers different factors in the objective function, including the workers' labels, workers' reliability, the confidence in the estimation, the machine learning predictions, and the characteristics of sequential labeling tasks. The proposed solution iteratively updates the machine learning model, workers' weights, and aggregation results. $AggSLC$ handles complex sequential label aggregation problems with fewer parameters comparing the state-of-the-art and produces theoretical guaranteed results. Our experimental results on real-world and simulated datasets illustrate that $AggSLC$ outperforms the state-of-the-art sequential label aggregations methods.  

\section{Related Works}
\noindent Data aggregation and label inference tasks have gained lots of concentration over the past decade, and many methods are developed to handle various challenges \cite{li2016survey, zheng2017truth}. We summarize the related works into two categories as below.

\textbf{{Aggregation of Crowdsourced Annotations.}} The maximum likelihood estimation method (D\&S) \cite{Dawid1979MaximumLE} aims to model the annotator's reliability using confusion matrices and infer the hidden truth labels through maximum likelihood. 
Many follow-up methods \cite{YHY08, 2008cheap, whitehill2009whose, groot2011learning} are inspired by the D\&S model. 
Later, optimization-based methods are proposed \cite{zhou2012learning, CRH14, li2014confidence}.  
Zhou et al. \cite{zhou2012learning} propose a minimax entropy principle for estimating the true labels from the annotations provided by crowds. 
Li et al. \cite{CRH14} model the problem as minimizing the overall weighted deviation between the truths and the multi-source observations where each source is weighted by its reliability. In the aforementioned methods, the instances are assumed to be independent. 
More recently, methods have been developed to handle various types of correlations among annotation instances, such as spatial-temporal dependencies among instances \cite{meng2016tackling, yao2018online, zhi2018dynamic}. Recent survey papers \cite{li2016survey, zheng2017truth} have shown that the aggregation methods which consider the worker reliabilities significantly outperform the naive aggregation methods such as voting.

More relevant to this paper, several methods \cite{rodrigues2014sequence, nguyen2017aggregating, simpson-gurevych-2019-bayesian, sabetpour2020optsla} are proposed to handle the sequential labeling tasks in NLP. Rodrigues et al. \cite{rodrigues2014sequence} proposed a probabilistic approach using Conditional Random Fields (CRF) to model the sequential annotations. In this model, the worker's reliability is modeled by their F1 score, but only one worker is assumed to be correct for any instance. Nguyen et al. relaxed the assumption and proposed a hidden Markov model (HMM) extension \cite{nguyen2017aggregating}. 
Recently, Simpson et al. \cite{simpson-gurevych-2019-bayesian} proposed a fully Bayesian approach, using a $J\times J\times J$ tensor to model each worker's reliability, where $J$ is the number of classes in the labeling task. 
\begin{figure*}[htb]
\centering
 \subfloat [Traditional truth discovery for sequence labeling]{
\includegraphics[width=0.3\textwidth]{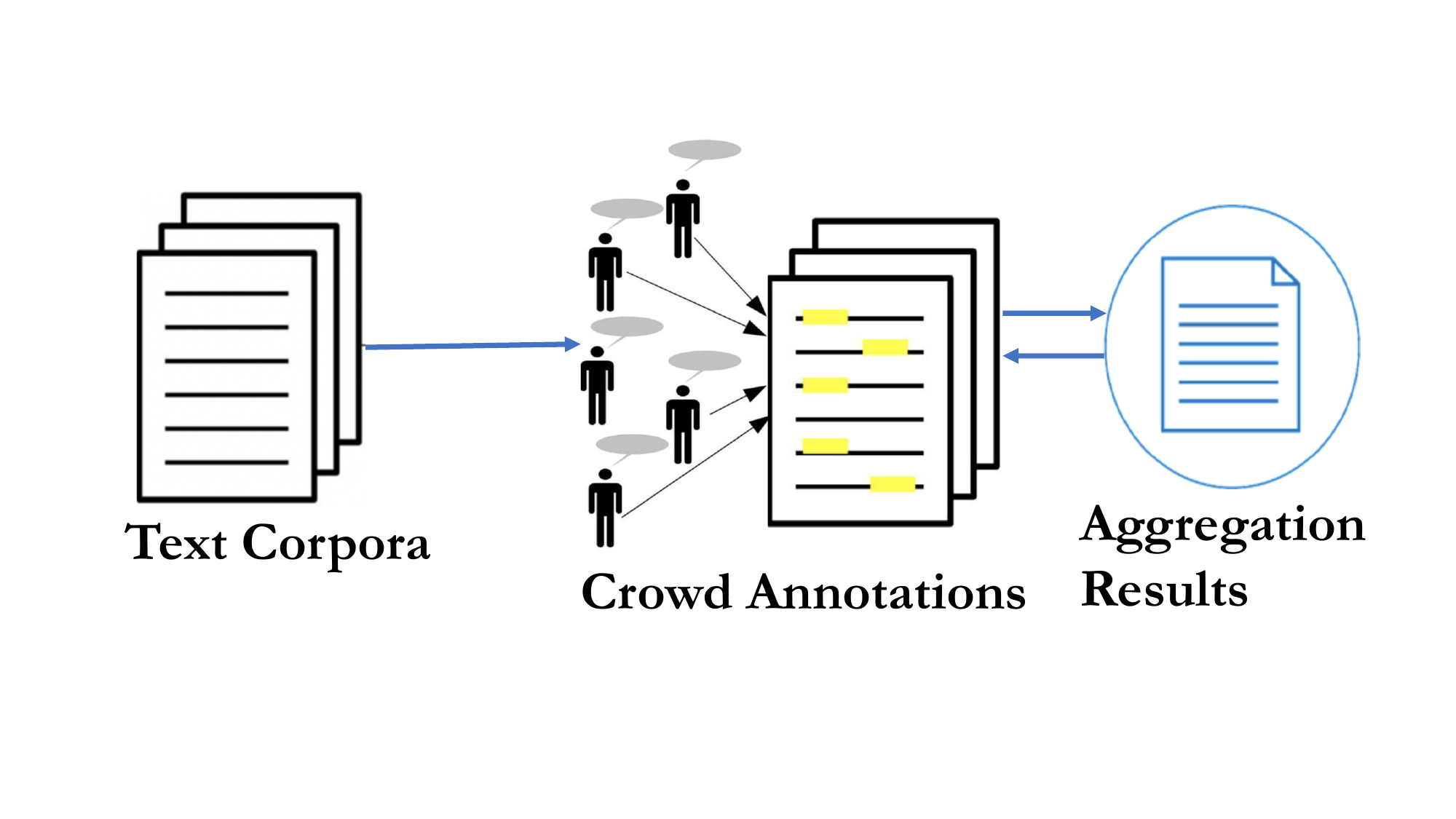}

 }
\hfil
 \subfloat [Deep learning with noisy labels]{
\includegraphics[width=0.3\textwidth]{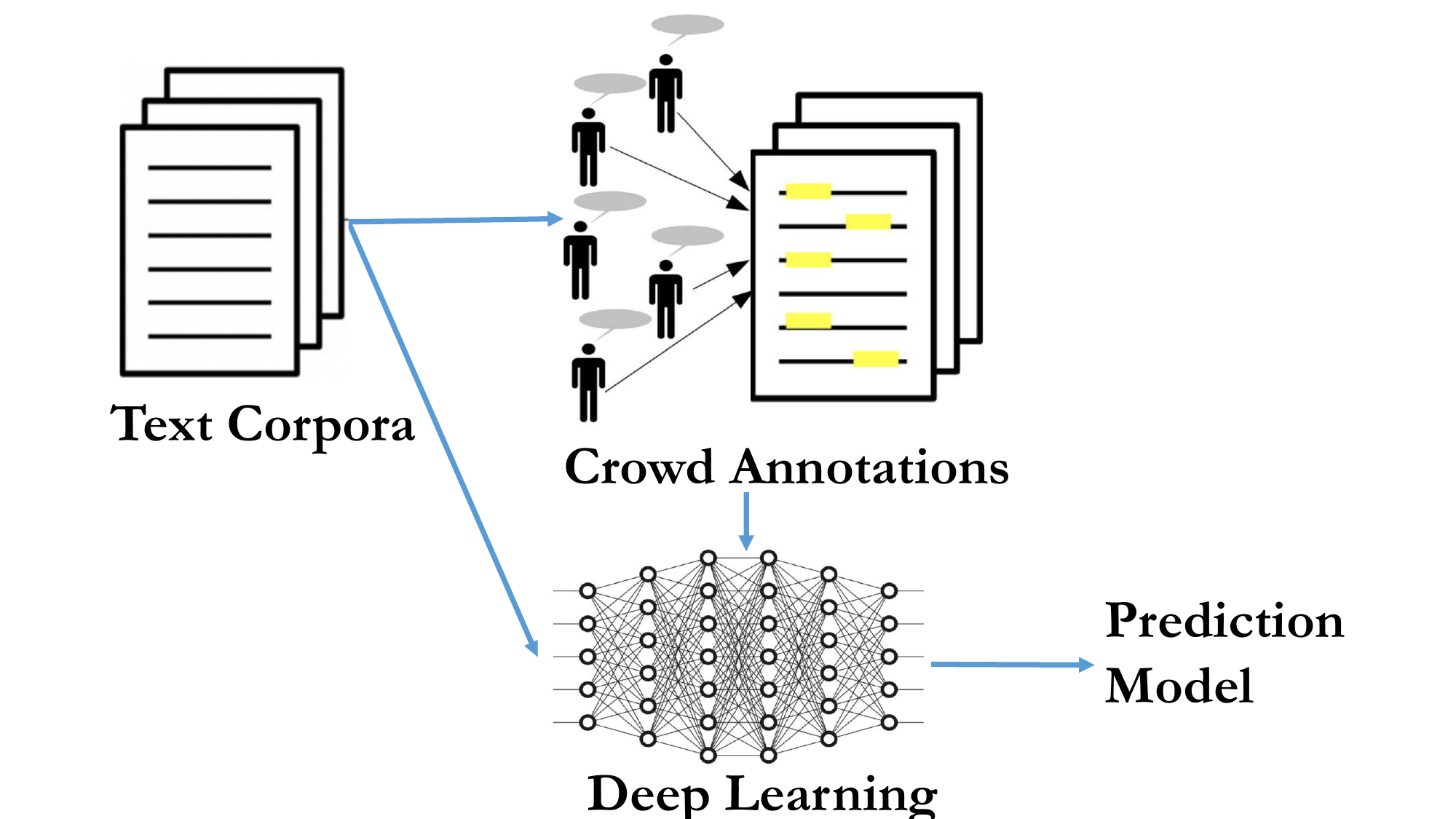}

 }
\hfil
\subfloat [$AggSLC$ framework]{
\includegraphics[width=0.3\textwidth]{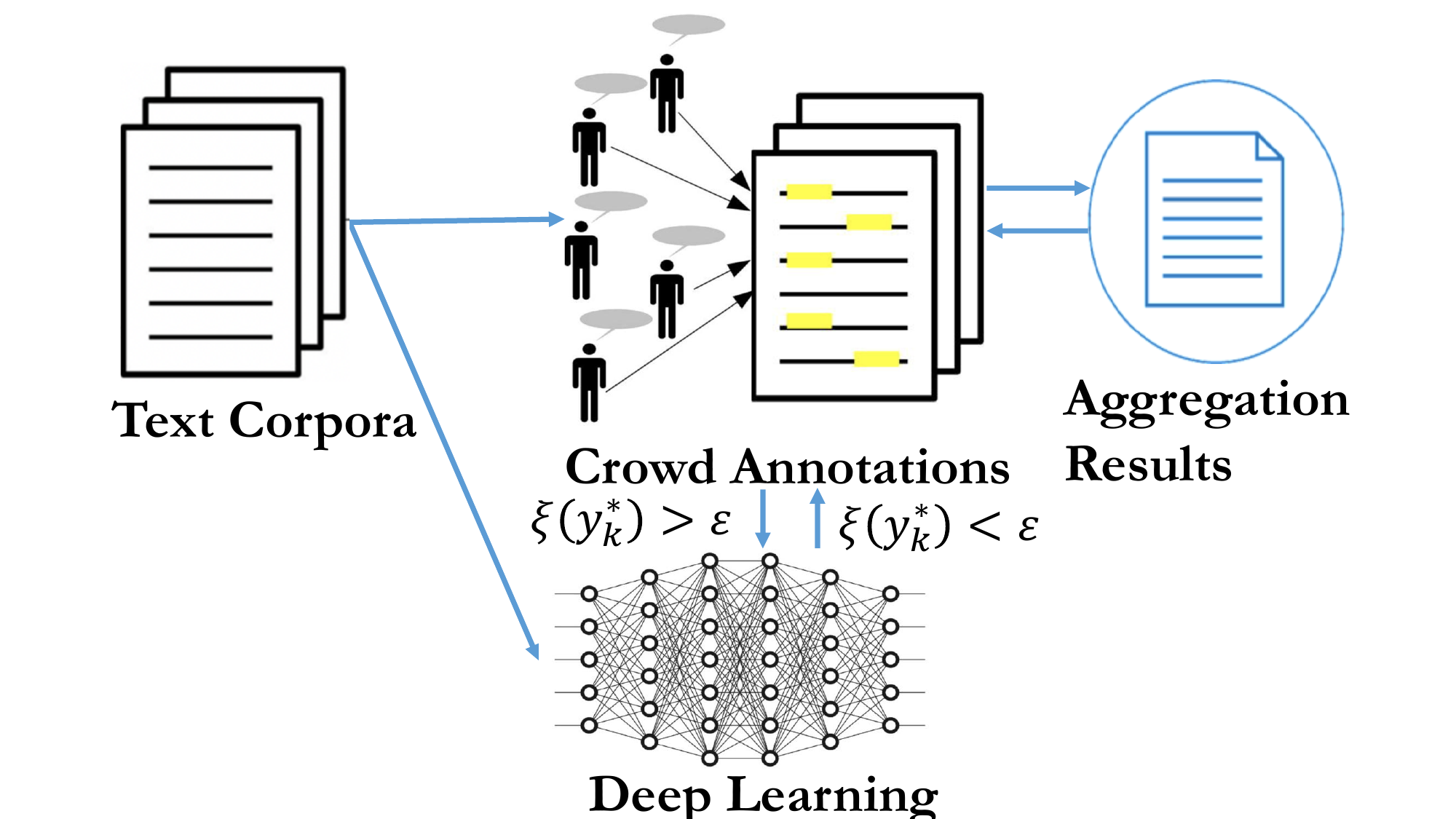}

 }
\vfil
\caption{Comparison of different lines of methods.}
\label{fig:comparison}
\end{figure*}
 
Recently, an optimization approach (OptSLA) has been proposed \cite{sabetpour2020optsla} to aggregate the sequential labels. Their approach to follow the sequential labeling rules is restricted by the unidirectional consistency loss function, which does not consider the consistency with the previous token label. None of the previous works provide theoretical guarantees. 
    
\textbf{{Deep learning with noisy labels.}} As noisy labels critically deprave deep neural networks' generalization performance, robust training and learning from noisy labels have received more and more attention in modern deep learning applications \cite{song2020learning, frenay2013classification, zhang2016learning}. 
For example, robust loss
functions and loss adjustment \cite{ghosh2017robust, zhang2018generalized, wang2019symmetric, lyu2019curriculum}, robust deep neural network architectures \cite{chen2015webly, sukhbaatar2014training, han2018masking, yao2018deep}, and robust regularization techniques \cite{goodfellow2014explaining, pereyra2017regularizing, tanno2019learning} are proposed to handle the noise in training data.

 In another line of research, the EM algorithm is used to jointly estimate the parameters of the machine learning model and the reliabilities of the workers \cite{JMLR:v11:raykar10a,albarqouni2016aggnet}. However, for these EM-based methods, the computational complexity of training is high due to the complex iterative procedure. Therefore, multiple methods \cite{guan2017said, rodrigues2018deep} improve the efficiency by training the deep neural networks directly from the noisy labels. 
 Unlike deep learning methods with noisy labels, which aim to establish prediction models, $AggSLC$ is an aggregation model seeking to combine the annotations from various sources. Inspired by the deep learning models, our framework's prediction module is designed to boost aggregation performance. An illustrative comparison between the existing branches of studies with the proposed method framework is displayed in Figure \ref{fig:comparison}.
   
\section{Methodology}

\noindent We formulate the problem of sequential label aggregation as an optimization problem and incorporate various loss functions. The proposed aggregation model formulates the annotation aggregation problem using a weighted loss of the inferred sequence of aggregated labels to the reliable workers' labels and machine learning predictions. $AggSLC$ further contains the inconsistency loss function to follow the sequential labeling rules. 
\subsection{Problem Formulation} 

\noindent We first introduce the notations frequently used in this paper (summarized in Table \ref{Notation}). Let us assume that $m$ workers (indexed by $j$) are hired to annotate the corpus, which consists of $s$ sentences (indexed by $k$) each containing $n$ tokens (indexed by $i$). $y_{i_k}^j$ is a one-hot vector that denotes the annotation given by the $j$-th worker on the $i$-th token in the $k$-th sentence. We use $\hat{y}_{i_k}$ to denote the machine learning prediction on this token. Each worker has a weight parameter $w_j$ to reflect his/her annotation quality, and $\mathcal{W}=\{w_1,w_2,...,w_m\}$ refers to the set of all worker weights. A higher weight implies that the worker is of higher reliability. We use $w_{dl}$ to refer to the weight of machine learning.

The goal is to infer the best sequence of aggregated annotations ($\mathbf{Y}^*_k$) that have the minimum disagreement with the sequence of annotations provided by workers and the machine learning predictions and follows the sequential labeling rules. 
\begin{table}
\centering
\caption{Summary of Notations}
\label{Notation} 
\begin{tabular}{ccl} 
\toprule
    Notation & Definition\\
\midrule
\ $m$ & number of workers indexed by $j$\\ [0.1cm]
\ $n$ & number of tokens indexed by $i$\\ [0.1cm]
\ $s$ & number of sentences indexed by $k$\\ [0.1cm]
\ $i_k$ & the $i$-th token in the $k$-th sentence\\ [0.1cm]
\ $y_{i_k}^j$ & workers' labels for token $i_k$ \\ [0.1cm]
\ $\mathbf{Y}_k^j$ & sequence of workers' labels for a sentence \\ [0.1cm]
\ $y_{i_k}^\ast$ & aggregated annotation for token $i_k$ \\ [0.1cm]
\ $\mathbf{Y}_k^\ast$ & sequence of aggregated annotations for a sentence \\ [0.1cm] 
\ $\hat{y}_{i_k}$ & machine learning prediction for token $i_k$\\ [0.1cm]
\ $\hat{\mathbf{Y}}_k$ & sequence of machine learning prediction for a sentence \\ [0.1cm]
\ $w_{j}$ & the weight of the $j$-th worker \\ [0.1cm]
\ $\mathcal{W}$ & the set of workers' weights $\{w_1,w_2,...,w_m\}$\\
\ $w_{dl}$ & the weight of the machine learning predictions\\ [0.1cm]
\ $\boldsymbol{\theta}$ & machine learning parameters\\
\bottomrule
\end{tabular}
\end{table}
\subsection{The proposed framework}
\noindent In this section, we introduce the $AggSLC$ framework and define each term in detail. 
The basic idea of the proposed framework is that the labels are mostly to be correct if they are provided by reliable workers, agreed by the high-quality machine learning method, and consistent with the sequential labeling rules. Therefore, our goal is to minimize the overall weighted loss of the aggregated sequential annotations $\mathbf{Y}^*_k$ to the sequence of labels provided by reliable workers $\mathbf{Y}_{k}^j$, machine learning predictions for the sentence ${\hat{\mathbf{Y}}_k}$, and the loss of inconsistencies in sequence labeling. 

Mathematically, we formulate the aggregation problem as an optimization problem with respect to the aggregated annotation $\mathbf{Y}^*$, a set of worker weights $\mathcal{W}$, the weight of machine learning model $w_{dl}$, and the machine learning parameters $\mathbf{\boldsymbol\theta}$ shown in Eq. (\ref{eq 1}):  
\begin{align} \label{eq 1}
&\min_{{\mathbf{Y}^*, \mathcal{W}, w_{dl}, \boldsymbol\theta}} \mathcal{F}=\min_{{\mathbf{Y}^* , \mathcal{W}, w_{dl}, \boldsymbol\theta}} (L_{agg} + L_{pred} + L_{inc}) \nonumber \\
&s.t. (\sum_{j=1}^m exp(-w_j))+ exp(-w_{dl})=1. 
\end{align}
The objective function $\mathcal{F}$ consists of three components. $L_{agg}$ refers to the loss from the aggregation module, $L_{pred}$ refers to the loss from the prediction module, and $L_{inc}$ indicates the loss from the consistency module. The regularization function is defined to guarantee the weights are always non-zero and positive. 

The aggregation module $L_{agg}$ computes the dissimilarity between inferred annotations to workers' labels (discussed in Section \ref{sec:agg}). The prediction module $L_{pred}$ calculates the dissimilarity between inferred annotations to machine learning prediction (discussed in Section \ref{sec:pred}). The consistency module $L_{inc}$ aims to maintain the label consistency between tokens for sequence labeling tasks (discussed in Section \ref{sec:inc}).

\subsubsection{\textbf{Aggregation module}} \label{sec:agg}
The aggregation module aims to find the aggregated annotation based on the workers' labels. We follow the same principle proposed in \cite{CRH14}. The idea is that the loss is significant if the estimated true labels are far from the labels given by the workers of high weights. By minimizing the loss, the aggregated labels will be close to the annotations from reliable workers. A reliable worker will get a high weight if their annotations are close to the aggregated labels. Mathematically, the weighted loss is formulated in Eq. (\ref{Eq 2}) as follows:
\begin{align} \label{Eq 2}
   L_{agg}(\mathcal{W}, \mathbf{Y}^*, \mathbf{Y}^j) = \sum_{j}w_j\sum_{k}\xi(\mathbf{Y}_k^\ast)  \sum_{{i_k}}H(y_{i_k}^j,y_{i_k}^\ast), 
\end{align}
\indent where $\xi(\mathbf{Y}_k^\ast)$ refers to the confidence score of the annotations, and function $H(\cdot, \cdot)$ is the cross-entropy loss to measure the disagreement level between two vectors.

The loss is further adjusted by the confidence score $\xi(\mathbf{Y}_k^\ast)$. 
Intuitively, if the annotations are agreed by most workers, we have higher confidence in the aggregation results. For simplicity, we define the confidence score on the sentence level instead of the token level to establish the machine learning prediction in the future steps. For sentences with higher confidence, if a worker's annotations are pretty different from the aggregation labels, they should receive a higher penalty. On the contrary, for sentences with lower confidence, they should get a lower penalty.  
The confidence score $\xi(\mathbf{Y}^*_k)$ is defined as:
\begin{equation} \label{Eq 4}
    \xi(\mathbf{Y}^*_k)=\frac{1}{l_k} \sum_{i_k}margin (y_{i_k}^*),
\end{equation}
\indent where $l_k$ is the number of tokens in sentence $k$ and $margin (y_{i_k}^*)$ is the probability difference between the two most likely labels of $y_{i_k}^*$.
If the same label is provided for a given token by most workers, the probability of that label being correct is higher. $\xi(\mathbf{Y}_{i_k}^*)$ considers agreement of the workers' labels for the given sentence. 
The higher agreement among workers' labels leads to a higher margin and thus a higher confidence score ($ \xi(\mathbf{Y}^*_k)$). On the other hand, if the workers disagree, the margin between the top two labels is lower, leading to a lower confidence score.

\subsubsection {\textbf{Prediction module}} \label{sec:pred}
The prediction module aims to obtain the annotations from a trained machine learning model. With deep learning, many state-of-the-art machine learning models can achieve high-performance \cite{devlin2018bert}. In our experiments, we treat the machine learning prediction as an additional worker. The loss is formulated similarly as follows:
\begin{equation} \label{Eq 5}
  L_{pred}(w_{dl} , \mathbf{Y}^*, \hat{\mathbf{Y}})= w_{dl}\sum_{k}\xi(\mathbf{Y}_k^\ast)\sum_{i_k}H(\hat{y}_{i_k},y_{i_k}^\ast). 
 \end{equation}
\indent The weighted cross-entropy loss between $y_{i_k}^*$ and the predicted labels $\hat y_{i_k}$ from a trained machine learning model is calculated. The weight of the machine learning model $w_{dl}$ captures the reliability of the machine learning prediction. The loss is adjusted by the confidence score calculated in the aggregation module.
The challenge of the prediction module is to obtain a high-quality machine learning model. The training of the machine learning model is discussed in Section \ref{Sec:DL}.

\subsubsection{\textbf{Consistency module}} \label{sec:inc}
The consistency module aims to obtain coherent labels for the sequential labeling task. The term $L_{inc}$ as shown in Eq. (\ref{Eq 6}) is a loss function that maintains the sequential label consistency in the inferred aggregation labels. 
It is beneficial to access both the left and right context to keep the consistency between two successive tokens according to the sequential label rules.
We formulate the bidirectional consistency module considering both previous and succeeding tokens' labels in the sentence as shown in Eq. (\ref{Eq 6}):
\begin{align} \label{Eq 6}
    L_{inc}= \sum_k\sum_{i_k} (l_{inc}(y_{{i_k}-1}^\ast,  y_{i_k}^\ast)+l_{inc}(y_{{i_k}}^\ast, y_{{i_k}+1}^\ast)), 
\end{align}

where function $l_{inc}(\cdot,\cdot)$ refers to the consistency loss between two consecutive token labels in aggregated sequential annotations as defined in Eq. (\ref{Eq inc}):
\begin{equation} \label{Eq inc}
  l_{inc}(y_{i_k}^*,y_{{i_k}+1}^* )=\begin{cases}
    0, & \text{if $P(y_{{i_k}+1}^*|y_{i_k}^*)>0  $}.\\
    1, & \text{$Otherwise$}.
  \end{cases}.
\end{equation}

This function will give $0$ loss if $P(y_{{i_k}+1}^*|y_{i_k}^*)$ is positive meaning the corresponding label sequence of $y_{i_k}^\ast,y_{{i_k}+1}^\ast$ is valid according to sequential label rules, and $1$ if the sequence is invalid. $l_{inc}(y_{i_k-1}^*,y_{{i_k}}^* )$ is calculated similarly. 

\section{The joint optimization problem}
To optimize the objective function Eq. (\ref{eq 1}), we apply the gradient descent algorithm \cite{tseng2001convergence} by iteratively updating the model parameters. We start with majority voting as an initial estimate of aggregation and then iteratively conduct the workers' weight update and aggregation update steps until convergence.

\subsection{Weights Update} 
To update the weights in the model, we treat the other variables as fixed. Then the weight update is calculated using Eq. (\ref{Eq 9}) as
\begin{align} \label{Eq 9} 
   \mathcal{W}\xleftarrow\ \operatorname*{argmin}_\mathcal{W} \mathcal{F}(\mathbf{Y}^* , \mathcal{W}, w_{dl} , \mathbf{\boldsymbol\theta}).
\end{align}
\indent The closed-form solution can be obtained by taking a partial derivative of Eq. (\ref{eq 1}) with respect to workers' weights and the machine learning weight. The solution for worker weight update is as follows:
\begin{equation} \label{Eq 10} 
   w_j= -\log \frac {\sum_{k}\xi(\mathbf{Y}_k^\ast) \sum_{i_k} H(y_{i_k}^j,y_{i_k}^*)}{\max_{k}\sum_{k}\xi(\mathbf{Y}_k^\ast)\sum_{i_k} H(y_{i_k}^j,y_{i_k}^\ast)} .  
\end{equation}
\indent The machine learning weight $w_{dl}$ is updated similarly.
The total worker annotation loss is the sum of the cross-entropy loss calculated between the aggregated labels and the labels provided by the worker over all the annotated sentences. 
The above equation shows that the worker weight is inversely proportional to the maximum annotation loss between workers. We normalize workers' annotation loss as different workers can work on other numbers of sentences.
If a worker's annotations are close to the aggregated labels, they will get a high weight, indicating that this worker is reliable.
\subsection{Aggregated Annotation Update }
With the updated workers' weight and machine learning weight, the aggregated annotation update ($\mathbf{Y}^*$) to minimize Eq. (\ref{eq 1}) is conducted as follows:
 \begin{align} \label{Eq 12}
      {\mathbf{Y}^*}\xleftarrow \ 
      &\operatorname*{argmin}_{\mathbf{Y}^*}  
     (\sum_{j}w_j\sum_{k} \xi(\mathbf{Y}_k^\ast)  \sum_{i_k}H(y_{i_k}^j,y_{i_k}^\ast) \nonumber\\
+&w_{dl}\sum_{k}\xi(\mathbf{Y}_k^\ast)\sum_{i_k}H(y_{i_k}^\ast,\hat{y}_{i_k}))\nonumber\\
+&\sum_k\sum_{i_k}(l_{inc}(y_{{i_k}-1}^\ast, y_{i_k}^\ast)  +l_{inc}(y_{{i_k}}^\ast, y_{{i_k}+1}^\ast)). 
\end{align} 

This equation does not have a closed-form solution. In fact, because of the variable correlation in the consistency loss function $l_{inc}(\cdot, \cdot)$, the solution is non-trivial. We apply the gradient descent to calculate $\mathbf{Y}^{1^*}$ ($1^*$ refers to the first intermediate step in updating the aggregated annotations) without the consistency loss function first, and then 
we introduce a \textit{bidirectional solution} inspired by the Viterbi algorithm \cite{nguyen2007comparisons} to incorporate the consistency loss function and calculate $\mathbf{Y}^*$. 

\noindent \textbf{The proposed bidirectional approach.}

\noindent Our proposed bidirectional solution includes two traversals: forward traversal, which aims to enforce the consistency between the current and the next inferred aggregated annotation ($l_{inc}(y_{i}^\ast, y_{i+1}^{\ast})$), and backward traversal which aims to enforce the consistency between the previous and the current inferred aggregated annotation ($l_{inc}(y_{i-1}^\ast, y_i^\ast)$). We start our solution with the forward traversal. Formally, given the sequence of aggregated annotations for the $k$-th sentence at step 1 ($\mathbf{Y}_k^{1^*}$), we minimize Eq. (\ref{Eq 12}) by minimizing the loss of inconsistency.

The loss for the sequence of token labels at step 1 ($\mathbf{Y}_k^{1^*}$) is the sum of loss for individual token labels. We aim to minimize the loss for the sequence to obtain the consistent sequence of aggregated annotations $\mathbf{Y}_k^{2^*}$ as shown in Eq. (\ref{Y-2}):
\begin{equation} \label{Y-2}
   \mathbf{Y}_k^{2^*} \xleftarrow\ \operatorname*{argmin} (\sum_{i_k} l_{inc} (y_{i_k}^{1^*},y_{i_k+1}^{1^*})).
\end{equation}

To minimize Eq. (\ref{Y-2}), we apply the Viterbi algorithm by defining the observation space, transition probability, and emission probabilities as follows.
The observation space is all the possible labels in the dataset. The transition probability between two consecutive labels is 1 if $l_{inc}(\cdot,\cdot)=0$, and 0 if $l_{inc}(\cdot,\cdot)=1$. The emission probability for $y_{i_k}^{1^*}$ equals to 1, if $l_{inc}(\cdot,\cdot)=0$, and if $l_{inc}(\cdot,\cdot)=1$ then the emission probability is defined as the probability of selecting a label from observation space for a token given worker labels. The most probable sequence of labels ($\mathbf{Y}_k^{2^*}$) obtained from the Viterbi algorithm ensures that Eq. (\ref{Y-2}) is minimized.

The consistent sequence of aggregated annotations for backward traversal ($\mathbf{Y}_k^{3^*}$) is shown in Eq. (\ref{Y-3}):
\begin{equation} \label{Y-3}
   \mathbf{Y}_k^{3^*} \xleftarrow\ \operatorname*{argmin} (\sum_{i_k} l_{inc} (y_{i_k-1}^{1^*},y_{i_k}^{1^*})),
\end{equation}
and is computed similarly with the forward traversal.

Both the two sequences of aggregated annotations $\mathbf{Y}_k^{2^*}$ and $\mathbf{Y}_k^{3^*}$ are consistent with the sequential labeling rules. We choose the sequence with higher recognized entity labels as the final sequence of aggregated annotation $\mathbf{Y}_k^*$. If both $\mathbf{Y}_k^{2^*}$ and $\mathbf{Y}_k^{3^*}$ have the same number of detected entity labels, we randomly pick one of them.

\subsection{Machine Learning Model Update}
\noindent We propose to simultaneously train a machine learning model along with the aggregation process. There are two major challenges: First, the aggregated labels are still noisy. If a machine learning model is trained on noisy labels, it may not perform well. Second, the trained machine learning model is supposed to provide predictions to the corpus in the prediction module. If a model is trained on the same corpus with the noisy labels, the model can overfit the noise in training data, thus being unable to provide useful predictions. 

To tackle these two challenges, we propose to use sentences with high confidence scores $\xi(\mathbf{Y}_k^*)$ (e.g., $\xi(\mathbf{Y}_k^*) > 0.9)$ for training, instead of using the entire corpus with noisy labels. The labels for these sentences are most likely correct, so the training data contain less noise. The trained model can then predict annotations to sentences with a lower confidence score (e.g., $\xi(\mathbf{Y}_k^*) \leq 0.9)$ in the prediction module. The Bidirectional Encoder Representations from Transformers (BERT) \cite{devlin2018bert} is employed as the deep learning model in our experiments.
The pseudo-code of $AggSLC$ is summarized in Algorithm \ref{Algo 1}.
\begin{algorithm}
\caption{The proposed Framework}\label{Algo 1}
\begin{flushleft}
\emph{\textbf{Input}: $s$ sentences indexed by $k$, workers annotations \{$y_{i_k}^1$,$y_{i_k}^2$,...,$y_{i_k}^j$}\}

\emph{\textbf{Output}: aggregated annotations $\mathbf{Y}_k^*$, workers' weights $w_j$}, machine learning weight $w_{dl}$, and a trained machine learning model 
\end{flushleft}
\begin{flushleft}

Initialize the aggregated annotation by $MV$ \\
\While {not convergence}{
 Update workers' weights $\mathcal{W}$ according to Eq. (\ref{Eq 10})\;
 
 Incrementally train the machine learning model with sentences having high confidence scores $\xi(\{y_{k}^*\})$\;
 
 Obtain predictions $\hat{y}_{i_k}$ from trained machine learning model\;

 Update $w_{dl}$ in the same way as workers' weight update following Eq. (\ref{Eq 10})\;
 
 \For {each sentence}{
    {Update $\mathbf{Y}_k^*$ according to Eq. (\ref{Eq 12})
}}
}
\end{flushleft}

\end{algorithm}
\section{Discussion}\label{Sec:Disscussion}
\noindent Here, we mark several issues to make the framework more practical, including the efficiency improvement of the machine learning models and the class imbalance problem.

\textbf{Incremental Deep Learning.}\label{Sec:DL} 
 In the prediction module of $AggSLC$, as the learning model is only trained with the highly confident sentences, the training data increments after each iteration. To speed up the training process, we use the incremental learning technique \cite{lomonaco2016comparing} to train the machine learning model. To retain the knowledge learned and incrementally incorporate it without compromising what it has learned before in a given iteration, we store the model to adjust the model parameters in the future iterations when new training sentences are added.   

\textbf{Class Weights ($U$).} In many sequential labeling tasks, the datasets may have one dominant label causing label imbalance problem \cite{johnson2019survey}. Hence the models using such skewed data tend to display weak performance for low-frequency classes.
For example, in the NER dataset (discussed in Section \ref{sec : real-world-datasets}), which is used in our experiments, the label "O" dominates the entity annotations. To handle this problem, class weights ($U$'s) can be used to re-weight the classes. A higher $U$ will increase the weight for entity labels when calculating $\mathbf{Y}^*$. In our experiments, we adopt a common class weight calculation, where class weight is inversely proportional to class frequency in the dataset \cite{scikit-learn}.

\section {Theoretical Analysis}
\noindent In this section, we prove the convergence of Algorithm \ref{Algo 1}. First, we show that the proposed method is an instance of the EM algorithm \cite{moon1996expectation}. For this purpose, we use the Expectation-Maximization (EM) theory for Normal/Independent random variables (N/I) or Gaussian scale mixtures (GSMs) \cite{lange1993normal} and show that it is an instance of the EM algorithm for constrained maximum likelihood estimation under a N/I assumption \cite{ba2013convergence}.
We define $U$ as a positive variable representing the class weight for a given label which is discussed in Section \ref{Sec:Disscussion}, $P_U(u)$ is the probability distribution function of $U$, and $y_{i_k}^j$ as a M-variate one-hot vector. Let $\mu$ be a constant M-dimensional vector to adjust worker biases. The N/I random vector ${y_{i_k}^\ast}$ for any constant M-dimensional vector $\mu$, is defined as Eq. (\ref{Eq 13}):
\begin{align} \label{Eq 13}
    {y_{i_k}^\ast}=\mu + U^{-1/2}{y_{i_k}^j}.
\end{align}
The density of ${y_{i_k}^\ast}$ is given by Eq. (\ref{Eq 14}):\\
{\small \begin{align} \label{Eq 14} 
    P_{y^*_{i_k}} = \frac{1}{(2\pi)^{M/2}|\Sigma|^{1/2}}\exp\left(\frac{-1}{2}\kappa((y_{i_k}^j-\mu)^T{\Sigma}^{-1}(y_{i_k}^j-\mu))\right),
\end{align}} 
\noindent where $\Sigma$ is the covariance of the M-variate normal random vector $y_{i_k}^j$. $\kappa(x)$ is defined by Eq. (\ref{Eq (15)}) for all positive $x$.
\begin{align}\label{Eq (15)}
    \kappa(x)=-2\ln\left(\int_{0}^\infty u^{M/2}e^{-ux/2}P_U(u)du\right).
\end{align}

Eq. (\ref{Eq 14}) represents the density of an elliptically-symmetric random vector $y_{i_k}^\ast$, and Eq. (\ref{Eq (15)}) gives a canonical form of the function $\kappa(\cdot)$ that arises from a given N/I distribution \cite{ba2013convergence}. 

\noindent \textbf {{EM Algorithm.}} We assume $w_j$ to be an unknown vector representing worker weight and $z$ as the total number of workers who worked on the specific sentence. From multi-variate N/I vector $P_{y^*_{i_k}}$ with mean $\mu$, covariance $\Sigma$ respectively and $t=1,2,\dotsc,z$. Eq. (\ref{Eq 16}) defines the log-likelihood of $z$ samples.
\begin{align} \label{Eq 16}
        L({\{y_{i_k}^*\}}_{t=1}^z;w_j)=\frac{-1}{2}\sum_{t=1}^z\{{\kappa(\delta_{t}^2(w_j))+\ln|\sum_t(w_j)|\}}.
\end{align}

EM algorithm maximizes the $Q$ function as shown in Eq. (\ref{Eq 17}). Here $w_j^{(\ell)}$ shows the current estimation of ${w_j}$.
\begin{align} \label{Eq 17}
    Q(w_j|w_j^{(\ell)})=\frac{-1}{2}\sum_{t=1}^z\{\ \kappa'(\delta_t^2(w_j^{(\ell)})\delta_t^2(\theta) + \ln|\sum_t(w_j)| \}.
\end{align}
Maximizing the $Q$ function is equivalent to the original likelihood function. 

\noindent \textbf{Convergence of the proposed method as an instance of EM algorithm.} 

\noindent The EM algorithm generates a sequence of iterates ${\{x^{(\ell)}\}_{\ell=0}^\infty}$ so that the sequence of log-likelihoods ${\{L(x^{(\ell)})\}_{\ell=1}^\infty}$ is converged \cite{ba2013convergence}. We assume $\mathcal{S}$ to be a non-empty, closed, strictly convex subset of ${\mathbb R}^N$, and define M as Eq. (\ref{Eq 18}):
\begin{align} \label{Eq 18}
    M(y_{i_k}^*)=\textnormal{argmin}_{x\in \mathcal{S}}||x||_{y^*_{i_k}}^2.
\end{align}

To prove the convergence of Eq. (\ref{eq 1}), we first show that there is a link between cross-entropy loss function $H$ and $l_2$-norm, so that we can replace $l_2$-norm by cross-entropy loss function $H$ in Eq. (\ref{Eq 18}).
Taking the log of $l_2$-norm, we will obtain $\log(\|x\|)$ which is equivalent to the cross-entropy loss function \cite{zhang2018generalized}. Moreover, cross-entropy is a convex measurement like $l_2$-norm, and we know that both likelihood and log-likelihood have the same extremum. 

As convergence point is an extremum point for Eq. (\ref{eq 1}), where workers' weight does not update, substituting $l_2$-norm function with cross-entropy gives the same result. In Eq. (\ref{Eq 19}), the plugged in cross-entropy loss function into Eq. (\ref{Eq 18}) is demonstrated.
\begin{align} \label{Eq 19}
    M(y_{i_k}^*)=\textnormal{argmin}_{x\in \mathcal{S}}\log(||x||_{y^*_{i_k}}).
\end{align}

To prove the convergence of the iterations,
we assume $x^{(0)}\in \mathcal{S}$ and $\{x^{(\ell)}\}_{\ell=0}^\infty\in \mathcal{S}$ be a sequence such that $x^{(\ell+1)}=M(x^{\ell})$ $\forall$ $\ell$. We need to prove the following: (1) $\{x^{\ell}\}$ is
bounded and $\|x^{\ell}-x^{(x^{\ell+1})}\|\rightarrow 0$; (2) every limit point of $\{x^{(*)}\}_{\ell=1}^\infty$ is a fixed point of $M$; (3) every limit point of $\{x^{\ell}\}_{\ell=0}^\infty$ is a stationary point of the Eq. (\ref{eq 1}) over $\mathcal{S}$; and (4) the Eq. (\ref{eq 1}) converges monotonically to some stationary point $x^{(*)}$. We refer the reader to \cite{ba2013convergence} for detailed proof.

\section{Experiments}
\noindent In this section, to demonstrate the proposed method's effectiveness, we conduct experiments\footnote{Our code can be found at \url{https://github.com/NasimISU/Truth-Discovery-in-Sequence-Labels-from-Crowds}} on both real-world and simulated datasets.
\begin{table*} [!htbp]\small
\centering 
\caption{Performance comparison for the NER and PICO datasets. "S" refers to the strict and ``R'' refers to relaxed metrics.} 
\label{table: real-world}

\begin{adjustbox}{width=180mm,center}
\begin{tabular}{l| c c c c c c | c c c c c c}
\hline
& \multicolumn{6}{c|}{\text{NER}}   
& \multicolumn{6}{c}{\text{PICO}} 
\\

& \text{S-Prec.} &  \text{S-Rec.} & \text{S-F1} & \text{R-Prec.} &  \text{R-Rec.} & \text{R-F1} & \text{S-Prec.} &  \text{S-Rec.} & \text{S-F1} & \text{R-Prec.} &  \text{R-Rec.} & \text{R-F1}
\\
\hline 
MV &
79.9 & 55.3 & 65.4 & 93.17 & 81.98 & 85.02 & \textbf{65.21} & 43.23 & 51.99 & \textbf{94.42} & 89.42 & 89.60
\\
CRF-MA& 
80.29 & 51.20 &62.53 & - & - & - & - & - & - & - & - & 
\\ 
HMM-crowd & 
77.40 & 72.29 & 74.76 & 89.54 & 88.55  & 87.67 & 46.56 & 49.59 & 48.03 & 79.53 & 93.89 & 86.11
\\
BSC-seq & 
80.3 & 73.72 & 76.87 & \textbf{94.70}  & 86.40 & 90.36 & 55.56 & 51.70 & 53.56 & 84.53 & 94.12 & 89.07 
\\
OptSLA & 
79.42 & 77.59 & 78.49 & 91.31 & 92.01 & 90.54 & 64.61 & 49.17 & 55.84 & 92.28 & 93.76  & 90.84 
\\
\emph{$AggSLC$} & 
\textbf{83.02} & \textbf{78.69} & \textbf{80.79} & 92.64 & \textbf{92.47} & \textbf{91.63} & 64.03 & \textbf{52.62} & \textbf{57.77} & 92.20 & \textbf{95.15} & \textbf{93.65}
\\
\bottomrule 
\end{tabular}
\end{adjustbox}
\end{table*}
\subsection{Experiment Setup} \label{sec: Exp Setup}
\subsubsection{Evaluation Metrics} We use precision, recall, and F1 to evaluate the performance\footnote{We use the ground truth labels only for evaluation purposes.} under two settings:
strict and relaxed, defined as follows:

\textbf {Strict Metrics (S)}. The strict metrics measure the span level performance. The entire span must be matched to be considered as correct. There is no partial credit for a partial match. We refer to CoNLL 2003 metrics for span level precision, recall, and F1 scores\footnote{The conlleval evaluation function can be found at \url{http:// amilab.dei.uc.pt/fmpr/ma-crf.tar.gz.}}.

\textbf {Relaxed Metrics (R).} The relaxed metrics measure the token level performance. It relaxes the strict metrics by providing partial credit for an incomplete match\footnote{The evaluation function can be found at \url{https://scikit-learn.org/stable/modules/model_evaluation.html}}.

\subsubsection{Baseline Methods}
We compare the proposed model with the following state-of-the-art aggregation methods with crowdsourced labels.  

    \textbf{Majority Voting (MV).} This is a common approach to infer the aggregated annotations where the label with the higher number of votes is chosen.
    
    \textbf{CRF-MA.} This is a probabilistic approach for sequence labeling using Conditional Random Fields (CRF) \cite{rodrigues2014gaussian}. It uses an EM algorithm to jointly learn the CRF model, the annotator reliability, and the aggregated annotations. 
    
    \textbf{HMM-crowd.} This is an extension of HMMs \cite{1165342} and LSTMs \cite{lample2016neural}
    in which crowd component is introduced by including additional parameters for workers reliability.
    
    \textbf{BSC-seq.} This is a full Bayesian method for aggregating sequential labels \cite{simpson-gurevych-2019-bayesian}.
    
    \textbf{OptSLA.} This is a framework employing optimization for aggregating crowdsourced sequential labels \cite{sabetpour2020optsla}.

The results of HMM-crowd and BSC-seq are reproduced through their public repositories. The results of CRF-MA are reported from their original paper \cite{rodrigues2014sequence}.

\subsection{Experiments on Real-World Datasets}
In this section, we conduct experiments on two real-world datasets.

\subsubsection {Datasets} \label{sec : real-world-datasets}
We use two publicly available real-world datasets to test the effectiveness of the proposed method.

    \textbf{NER dataset}\footnote{Dataset can be found at \url{http://amilab.dei.uc.pt/fmpr/crf-ma-datasets.tar.gz}}.  
The original English portion of the CoNLL 2003 dataset \cite{sang2003introduction} includes over 21,000 annotated sentences from 1393 articles. This dataset is split into three sets: train, validation, and test \cite{sang2003introduction}. Rodrigues et al. \cite{rodrigues2014gaussian} crowdsourced labels for 400 articles in the train set. The selected sentences have annotations from 47 workers who identify the named entities and annotate them as Person, Location, Organization, or Miscellaneous. We use the crowdsourced labels collected by Rodrigues et al. \cite{rodrigues2014gaussian} for all aggregation methods. The evaluation is conducted on the same evaluation set proposed by Nguyen et al. \cite{nguyen2017aggregating}. 

    \textbf{PICO dataset}\footnote{Dataset can be found at \url{https://github.com/yinfeiy/PICO-data}}. This dataset is annotated for the Biomedical information extraction task \cite{nguyen2017aggregating}. It is annotated to recognize text spans, identifying the population registered in a clinical trial. It includes 5,000 medical paper abstracts describing randomized control trials (RCTs) annotated by Amazon Mechanical Turk workers, and among these, 200 abstracts are annotated by medical students as ground truth labels. The proposed method and all baselines are evaluated on the abstracts with the ground truth labels.

\subsubsection{Experimental Results} \label{real-world-results}
The results are summarized in Table \ref{table: real-world}. Comparing the strict metrics on the NER dataset and the PICO dataset, all methods perform better on the NER dataset. This is because PICO requires more domain knowledge in annotation, and thus the crowd workers provide much noisier labels. Comparing relaxed metrics on NER and PICO datasets, we observe higher performance gain on PICO dataset. This is because PICO dataset contains a fewer number of labels corresponding NER dataset. All the reported results are from the iteration at which the model converges\footnote{All the experiments are performed on the system with 2.6 GHz 6-Core Intel Core i7 processor and 16 GB memory.}. 

Among the baseline aggregation methods, MV achieves high precision but significantly lower recall, indicating that the workers behave conservatively when annotating the corpus. HMM-crowd and BSC-seq models, in general, improve the results compared with MV by considering the workers' reliabilities. However, comparing these two baselines, BSC-seq outperforms HMM-crowd, indicating that the full Bayesian model can better capture the transitions among sequential labels. The optimization-based method OptSLA further improves on HMM-crowd and BSC-seq, showing that optimization-based approaches can better capture workers' reliabilities compared to statistical methods. The proposed method $AggSLC$ further improves the results. It can be observed that $AggSLC$ outperforms the aggregation baselines on both the strict and the relaxed metrics in terms of F1 score on both datasets. 

\begin{figure}[t]
\centering
\includegraphics[width=65mm]{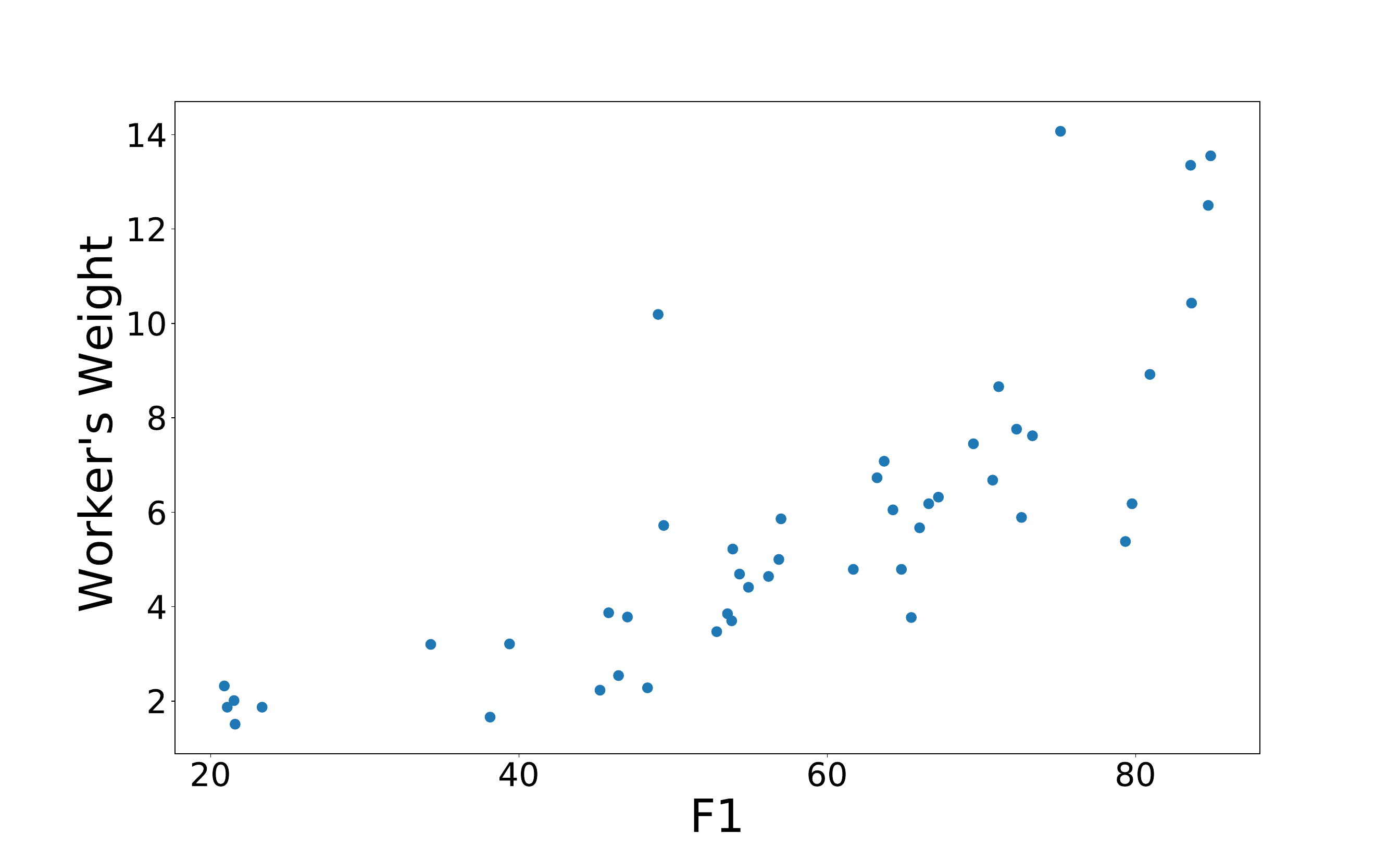}
\caption{Workers' weights w.r.t. their actual F1 score}
\label{fig:worker_weight}
\end{figure}

Furthermore, as the worker's reliability estimation is the key to obtain high-quality aggregation results, we further examine if the weights estimated by $AggSLC$ reflect the actual workers' reliabilities. We plot the estimated weights with respect to their actual F1 scores in Figure \ref{fig:worker_weight}. The workers are ranked based on their actual strict F1 scores. It can be observed that there is a strong positive correlation between worker weights and their actual F1 scores. We further calculate the Pearson correlation (which measures the linear correlation) and Spearman's rank coefficient (a non-parametric measure of rank correlation). For both measurements, a strong positive correlation between two variables will lead to a value close to $1$. The correlation coefficients between worker weights and their actual F1 scores are 0.79 and 0.87, respectively, illustrating a strong correlation. Moreover, because $AggSLC$ uses one parameter for each worker, the results are more straightforward to interpret comparing with the baseline methods.
\begin{figure}[t] 
  \centering
  \includegraphics[width=65mm]{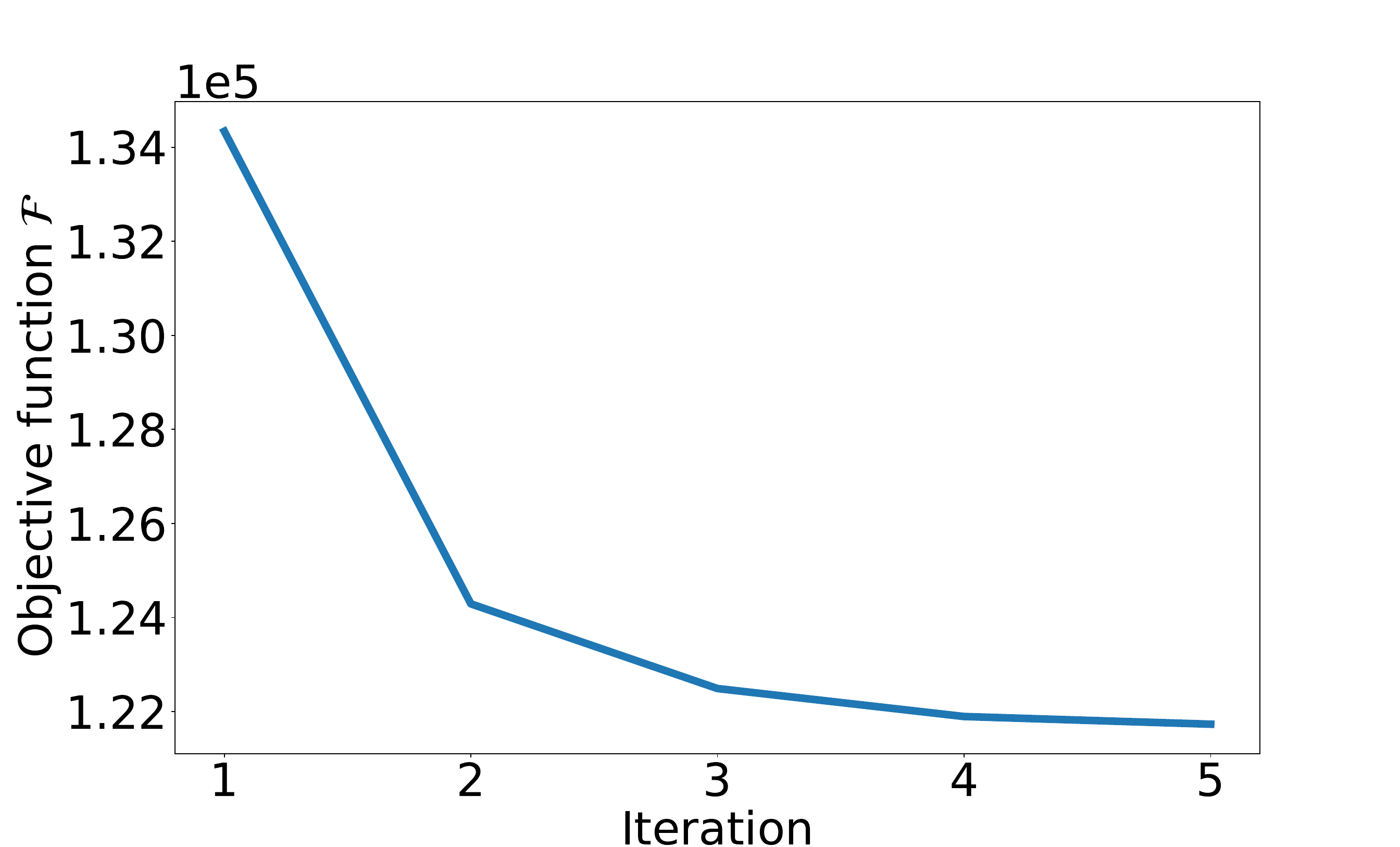}
  \caption{$AggSLC$ convergence on NER dataset}
  \label{fig:convergence}
\end{figure}
\subsubsection{Convergence Study.} In addition to the theoretical analysis on the convergence of $AggSLC$, we empirically examine the convergence. We use the NER dataset as the testbed for this experiment, and we observe similar results on the PICO dataset. Figure \ref{fig:convergence} illustrates the value of objective function ($\mathcal{F}$) with respect to each iteration. We can see that the value of the objective function decreases rapidly during the first three iterations and then reaches a stable stage, showing that the proposed method converges quickly in practice. As can be seen, ($\mathcal{F}$) is a decreasing function that converges to the minimal point.
\begin{table} [!htbp]
\centering 
\caption{Performance comparison for the synthetic dataset}
\label{table: simulated-data}
\begin{adjustbox}{width=90mm,center}
\begin{tabular}{l| c c c c c c}
\hline
& \multicolumn{6}{c}{\text{Synthetic Dataset}}\\
& \text{S-Prec.} &  \text{S-Rec.} & \text{S-F1} & \text{R-Prec.} &  \text{R-Rec.} & \text{R-F1}  \\
\hline 
MV &
79.03 & 39.41 & 52.6 & 95.85 & 70.68 & 77.81
\\
HMM-crowd & 
63.60 & 57.43 & 60.36 & 85.16 & 80.66 & 81.38
\\
BSC-seq & 
71.42 & 49.93 & 58.77 & 85.10 & 74.48 & 79.44
\\
OptSLA & 
\textbf{83.08} & 48.88 & 61.55 & \textbf{96.48} & 76.42 & 82.46
\\
\emph{$AggSLC$} & 
80.37 & \textbf{59.0} & \textbf{68.05} & 95.08 & \textbf{81.38} & \textbf{85.45}
\\
\bottomrule 
\end{tabular}
\end{adjustbox}
\end{table}
\begin{table*}
\centering
\caption{Effect of consistency module $L_{inc}$ and prediction module $L_{pred}$ on the aggregation performance} \centering 
\label{table: Ablation}
\begin{adjustbox}{width=180mm,center}
\begin{tabular}{l| c c c c c c | c c c c c c}
\toprule
& \multicolumn{6}{c|}{\text{NER}}   
& \multicolumn{6}{c}{\text{PICO}} \\
& \text{S-Prec.} &  \text{S-Rec.} & \text{S-F1} & \text{R-Prec.} &  \text{R-Rec.} & \text{R-F1} & \text{S-Prec.} &  \text{S-Rec.} & \text{S-F1} & \text{R-Prec.} &  \text{R-Rec.} & \text{R-F1}\\
\hline 
\small {$AggSLC$} & 
\textbf{83.02} & \textbf{78.69} & \textbf{80.79} & \textbf{92.64} & \textbf{92.47} & \textbf{91.63} & 64.03 & \textbf{52.62} & \textbf{57.77} & 92.20 & \textbf{95.15} & \textbf{93.65} 
\\
\small {$AggSLC$ - $L_{inc}$} & 
80.96 & 77.36 & 79.12  & 91.65 & 90.02 & 90.83 & 56.23 & 51.10 & 53.55 & 90.63 & 93.85 & 92.21 
\\
\small {$AggSLC$ - $L_{pred}$} & 
73.13 & 68.01 & 70.48  & 86.45 & 87.42 & 86.93 & \textbf{64.27} & 48.20 & 55.09 & \textbf{92.68} & 92.85 & 92.76 
\\
\small{$AggSLC$ - ($L_{inc}$,$L_{pred}$)} & 
72.47 & 64.62 & 68.32 & 85.58 & 86.16 & 85.87 & 56.07 & 48.48 & 52.00 & 90.18 & 93.26 & 91.69 
\\
\bottomrule
\end{tabular}
\end{adjustbox}
\end{table*}
\begin{table*} [!htbp] \small
\centering 
\caption{Performance comparison of the prediction module} 
\label{table: ablation2}

\begin{adjustbox}{width=180mm,center}
\begin{tabular}{l| c c c c c c | c c c c c c}
\hline
& \multicolumn{6}{c|}{\text{NER}}   
& \multicolumn{6}{c}{\text{PICO}} 
\\

& \text{S-Prec.} &  \text{S-Rec.} & \text{S-F1} & \text{R-Prec.} &  \text{R-Rec.} & \text{R-F1} & \text{S-Prec.} &  \text{S-Rec.} & \text{S-F1} & \text{R-Prec.} &  \text{R-Rec.} & \text{R-F1}
\\
\hline 
DL\_CL & 
66.00 & 59.30 & 62.40 & 82.45 & 79.79  & 78.80 & 25.73 & \textbf{47.65} & 33.41 & 77.71 & \textbf{92.10} & 79.73
\\
$L_{pred}$\ (ML model) & 
\textbf{70.95} & \textbf{77.16} & \textbf{73.93} & \textbf{84.01} & \textbf{91.31} & \textbf{87.51} & \textbf{38.9} & 46.69 & \textbf{42.44} & \textbf{78.51} & 91.73 & \textbf{84.60}
\\
\bottomrule 
\end{tabular}
\end{adjustbox}
\end{table*}

\subsection{Experiments on Synthetic Dataset} \label{synthetic-results}
\noindent We simulate a dataset using different knowledge bases and machine learning methods for the NER task to conduct more comprehensive experiments. We first introduce the simulation strategy to generate the dataset in Section \ref{dataset simulation}; next, we use the simulated dataset to evaluate the effectiveness of the proposed approach in Section \ref{sec:simul}.

\subsubsection{Dataset Simulation} \label{dataset simulation}
We simulate the workers' annotations using the sentences in the English portion of the CoNLL 2003 dataset introduced in Section \ref{sec : real-world-datasets}. The simulated dataset consists of six workers. To keep the diversity within labeling, three workers are simulated by the existing knowledge bases, and three are stimulated by the state-of-the-art NER methods. Following, we briefly explain each method: 

\textbf{1. Knowledge bases.} We used knowledge bases including Wikidata, DBpedia, and Schema employing pyspotlight package\cite{de1999spotlight} to simulate workers' behaviour\footnote{\url{https://pypi.org/project/pyspotlight/}}. 

\textbf{2. Machine Learning methods for NER task.}
We use SpaCy \cite{shelar2020named}, Apache OpenNLP \cite{dawar2019comparing}, and GATE \cite{cunningham2002gate} to generate the workers' annotations. These are all machine learning models adapted for many tasks, including named entity recognition. SpaCy utilizes a hybrid of HMM, decision tree analysis, and maximum entropy models. Apache OpenNLP includes perceptron-based machine learning and maximum entropy. General Architecture for Text Engineering (GATE) includes an information extraction system.

\subsubsection{Experimental Results}\label{sec:simul}
The experimental results comparing $AggSLC$ with different baselines on the simulated datasets are shown in Table \ref{table: simulated-data}. It can be observed that $AggSLC$ outperforms all baselines. The reason is that the baseline methods have strong prior assumptions about the worker's behavior learned from the existing datasets. Thus, they are less flexible to new datasets with different worker behavior. 

\subsection{Ablation Study}

\noindent To gain insights into our three-module framework, we investigate the effectiveness of components of $AggSLC$ via ablation study. We conduct three sets of experiments to investigate the importance of each module of $AggSLC$. $AggSLC$ - $L_{inc}$ means that the consistency module is removed from Eq. (\ref{eq 1}), $AggSLC$ - $L_{pred}$ means that the prediction module is removed from  Eq. (\ref{eq 1}), and $AggSLC$ - $(L_{inc}$,$L_{pred})$ means that only the aggregation module is kept in Eq. (\ref{eq 1}). Note that although we utilized BERT for the prediction module, $AggSLC$ can adopt any other model for the prediction module. The results are summarized in Table \ref{table: Ablation}.

\textbf{Effect of the consistency module ($L_{inc}$).} We evaluate the proposed model in the absence of $L_{inc}$ for both NER and PICO datasets. 

Comparing $AggSLC$ and $AggSLC$ - $L_{inc}$, we observe that strict F1 has decreased for both datasets in the absence of $L_{inc}$ confirming our expectation.   

\textbf{Effect of the prediction module ($L_{pred}$).} Similarly, we evaluate the proposed model in the absence of the prediction module $L_{pred}$ to show the effectiveness of this module. There are two sets of comparisons for this experiment: $AggSLC$ with $AggSLC$ - $L_{pred}$, and $AggSLC$ - $L_{inc}$ with $AggSLC$ - ($L_{inc}$,$L_{pred})$, where the difference for each pair is the absence of the prediction module. We observe a considerable drop in performance when the prediction module is removed. This is because the model misses one good worker since the prediction output is considered an extra set of worker labels. The decrease in the NER dataset is larger than the decrease in the PICO dataset because the NER dataset's machine learning performance is significantly better than machine learning performance on PICO dataset (shown in Table \ref{table: ablation2}). 

\textbf{Performance comparison of the prediction module ($L_{pred}$).} We further evaluate the performance of the prediction module used in $AggSLC$ compared to the DL\_CL \cite{rodrigues2018deep}. DL\_CL is a deep learning method employed on noisy crowdsourced labels that includes a crowd layer and trains the deep neural networks end-to-end using backpropagation. The results are summarized in Table \ref{table: ablation2}. The results of DL\_CL are reproduced through their public repository. Compared with DL\_CL, the model trained using $AggSLC$ achieves better F1 scores on both strict and relaxed metrics. It shows that although the machine learning model trained in $AggSLC$ uses smaller training data, it can still produce a good prediction model since the training data is of a lower noise level.
\section{Conclusions}
\noindent In this paper, we propose an innovative optimization-based approach $AggSLC$ for sequential label aggregation tasks. To obtain the best sequence of aggregated labels on imbalanced datasets with complex dependencies between tokens, our model jointly considers the workers' labels, workers' reliability, the machine learning predictions, and the characteristics of sequential labeling tasks in the objective function. Our experimental results on NER, PICO, and synthetic datasets illustrate that $AggSLC$ outperforms the state-of-the-art sequential label aggregations methods. We demonstrate that the trained deep learning model outperforms the state-of-the-art deep learning with crowdsourced labels method DL\_CL. We empirically and mathematically show the convergence of the proposed $AggSLC$. To mathematically prove the convergence, we also provide theoretical analysis. 
\section*{Acknowledgment}
\noindent We thank Dr. Kris De Brabanter from Iowa State University for his valuable suggestions. The work was supported in part by the National Science Foundation under Grant NSF IIS-2007941, NSF IIS-1909879, NSF CNS-1931042, and  NSF IIS-2008155.
Any opinions, findings, conclusions, or recommendations expressed in this document are those of the author(s) and should not be interpreted as the views of any U.S. Government.

\balance
\bibliographystyle{IEEEtranS}  
\bibliography{bibi}  

\begin{thebibliography}{10}
\providecommand{\url}[1]{#1}
\csname url@samestyle\endcsname
\providecommand{\newblock}{\relax}
\providecommand{\bibinfo}[2]{#2}
\providecommand{\BIBentrySTDinterwordspacing}{\spaceskip=0pt\relax}
\providecommand{\BIBentryALTinterwordstretchfactor}{4}
\providecommand{\BIBentryALTinterwordspacing}{\spaceskip=\fontdimen2\font plus
\BIBentryALTinterwordstretchfactor\fontdimen3\font minus
  \fontdimen4\font\relax}
\providecommand{\BIBforeignlanguage}[2]{{%
\expandafter\ifx\csname l@#1\endcsname\relax
\typeout{** WARNING: IEEEtranS.bst: No hyphenation pattern has been}%
\typeout{** loaded for the language `#1'. Using the pattern for}%
\typeout{** the default language instead.}%
\else
\language=\csname l@#1\endcsname
\fi
#2}}
\providecommand{\BIBdecl}{\relax}
\BIBdecl

\bibitem{albarqouni2016aggnet}
S.~Albarqouni, C.~Baur, F.~Achilles, V.~Belagiannis, S.~Demirci, and N.~Navab,
  ``Aggnet: deep learning from crowds for mitosis detection in breast cancer
  histology images,'' \emph{IEEE transactions on medical imaging}, vol.~35,
  no.~5, pp. 1313--1321, 2016.

\bibitem{ba2013convergence}
D.~Ba, B.~Babadi, P.~L. Purdon, and E.~N. Brown, ``Convergence and stability of
  iteratively re-weighted least squares algorithms,'' \emph{IEEE Transactions
  on Signal Processing}, vol.~62, no.~1, pp. 183--195, 2013.

\bibitem{chen2015webly}
X.~Chen and A.~Gupta, ``Webly supervised learning of convolutional networks,''
  in \emph{Proceedings of the IEEE International Conference on Computer
  Vision}, 2015, pp. 1431--1439.

\bibitem{cunningham2002gate}
H.~Cunningham, ``Gate, a general architecture for text engineering,''
  \emph{Computers and the Humanities}, vol.~36, no.~2, pp. 223--254, 2002.

\bibitem{dawar2019comparing}
K.~Dawar, A.~J. Samuel, and R.~Alvarado, ``Comparing topic modeling and named
  entity recognition techniques for the semantic indexing of a landscape
  architecture textbook,'' in \emph{2019 Systems and Information Engineering
  Design Symposium (SIEDS)}.\hskip 1em plus 0.5em minus 0.4em\relax IEEE, 2019,
  pp. 1--6.

\bibitem{Dawid1979MaximumLE}
A.~P. Dawid and A.~Skene, ``Maximum likelihood estimation of observer
  error-rates using the em algorithm.''

\bibitem{de1999spotlight}
G.~de~Haan, M.~Bekker, and H.~de~Greef, ``Spotlight,'' \emph{EACE Quarterly},
  vol.~3, no.~2, pp. 22--23, 1999.

\bibitem{devlin2018bert}
J.~Devlin, M.-W. Chang, K.~Lee, and K.~Toutanova, ``Bert: Pre-training of deep
  bidirectional transformers for language understanding,'' \emph{Proceedings of
  NAACL-HLT 2019, Association for Computational Linguistics}, p. 4171–4186,
  2019.

\bibitem{frenay2013classification}
B.~Fr{\'e}nay and M.~Verleysen, ``Classification in the presence of label
  noise: a survey,'' \emph{IEEE transactions on neural networks and learning
  systems}, vol.~25, no.~5, pp. 845--869, 2013.

\bibitem{ghosh2017robust}
A.~Ghosh, H.~Kumar, and P.~Sastry, ``Robust loss functions under label noise
  for deep neural networks,'' in \emph{Proceedings of the Thirty-First AAAI
  Conference on Artificial Intelligence}, 2017, pp. 1919--1925.

\bibitem{goodfellow2014explaining}
I.~J. Goodfellow, J.~Shlens, and C.~Szegedy, ``Explaining and harnessing
  adversarial examples,'' \emph{In International Conference on Learning
  Representations}, 2015.

\bibitem{groot2011learning}
P.~Groot, A.~Birlutiu, and T.~Heskes, ``Learning from multiple annotators with
  gaussian processes,'' in \emph{International Conference on Artificial Neural
  Networks}.\hskip 1em plus 0.5em minus 0.4em\relax Springer, 2011, pp.
  159--164.

\bibitem{guan2017said}
M.~Y. Guan, V.~Gulshan, A.~M. Dai, and G.~E. Hinton, ``Who said what: Modeling
  individual labelers improves classification,'' \emph{arXiv preprint
  arXiv:1703.08774}, 2017.

\bibitem{han2018masking}
B.~Han, J.~Yao, G.~Niu, M.~Zhou, I.~Tsang, Y.~Zhang, and M.~Sugiyama,
  ``Masking: A new perspective of noisy supervision,'' in \emph{Advances in
  Neural Information Processing Systems}, 2018, pp. 5836--5846.

\bibitem{johnson2019survey}
J.~M. Johnson and T.~M. Khoshgoftaar, ``Survey on deep learning with class
  imbalance,'' \emph{Journal of Big Data}, vol.~6, no.~1, p.~27, 2019.

\bibitem{lample2016neural}
G.~Lample, M.~Ballesteros, S.~Subramanian, K.~Kawakami, and C.~Dyer, ``Neural
  architectures for named entity recognition,'' \emph{arXiv preprint
  arXiv:1603.01360}, 2016.

\bibitem{lan2019learning}
O.~Lan, X.~Huang, B.~Y. Lin, H.~Jiang, L.~Liu, and X.~Ren, ``Learning to
  contextually aggregate multi-source supervision for sequence labeling,''
  \emph{arXiv preprint arXiv:1910.04289}, 2019.

\bibitem{lange1993normal}
K.~Lange and J.~S. Sinsheimer, ``Normal/independent distributions and their
  applications in robust regression,'' \emph{Journal of Computational and
  Graphical Statistics}, vol.~2, no.~2, pp. 175--198, 1993.

\bibitem{li2014confidence}
Q.~Li, Y.~Li, J.~Gao, L.~Su, B.~Zhao, M.~Demirbas, W.~Fan, and J.~Han, ``A
  confidence-aware approach for truth discovery on long-tail data,''
  \emph{Proceedings of the VLDB Endowment}, vol.~8, no.~4, pp. 425--436, 2014.

\bibitem{CRH14}
Q.~Li, Y.~Li, J.~Gao, B.~Zhao, W.~Fan, and J.~Han, ``Resolving conflicts in
  heterogeneous data by truth discovery and source reliability estimation,'' in
  \emph{Proc.\ of the ACM SIGMOD International Conference on Management of Data
  (SIGMOD'14)}, 2014, pp. 1187--1198.

\bibitem{li2016survey}
Y.~Li, J.~Gao, C.~Meng, Q.~Li, L.~Su, B.~Zhao, W.~Fan, and J.~Han, ``A survey
  on truth discovery,'' \emph{ACM Sigkdd Explorations Newsletter}, vol.~17,
  no.~2, pp. 1--16, 2016.

\bibitem{lomonaco2016comparing}
V.~Lomonaco and D.~Maltoni, ``Comparing incremental learning strategies for
  convolutional neural networks,'' in \emph{IAPR Workshop on Artificial Neural
  Networks in Pattern Recognition}.\hskip 1em plus 0.5em minus 0.4em\relax
  Springer, 2016, pp. 175--184.

\bibitem{lyu2019curriculum}
Y.~Lyu and I.~W. Tsang, ``Curriculum loss: Robust learning and generalization
  against label corruption,'' \emph{In International Conference on Learning
  Representations}, 2020.

\bibitem{meng2016tackling}
C.~Meng, H.~Xiao, L.~Su, and Y.~Cheng, ``Tackling the redundancy and sparsity
  in crowd sensing applications,'' in \emph{Proceedings of the 14th ACM
  Conference on Embedded Network Sensor Systems CD-ROM}, 2016, pp. 150--163.

\bibitem{moon1996expectation}
T.~K. Moon, ``The expectation-maximization algorithm,'' \emph{IEEE Signal
  processing magazine}, vol.~13, no.~6, pp. 47--60, 1996.

\bibitem{nguyen2017aggregating}
A.~T. Nguyen, B.~C. Wallace, J.~J. Li, A.~Nenkova, and M.~Lease, ``Aggregating
  and predicting sequence labels from crowd annotations,'' in \emph{Proceedings
  of the conference. Association for Computational Linguistics. Meeting}, vol.
  2017.\hskip 1em plus 0.5em minus 0.4em\relax NIH Public Access, 2017, p. 299.

\bibitem{nguyen2007comparisons}
N.~Nguyen and Y.~Guo, ``Comparisons of sequence labeling algorithms and
  extensions,'' in \emph{Proceedings of the 24th international conference on
  Machine learning}, 2007, pp. 681--688.

\bibitem{novotney2010cheap}
S.~Novotney and C.~Callison-Burch, ``Cheap, fast and good enough: Automatic
  speech recognition with non-expert transcription,'' in \emph{Human Language
  Technologies: The 2010 Annual Conference of the North American Chapter of the
  Association for Computational Linguistics}, 2010, pp. 207--215.

\bibitem{scikit-learn}
F.~Pedregosa, G.~Varoquaux, A.~Gramfort, V.~Michel, B.~Thirion, O.~Grisel,
  M.~Blondel, P.~Prettenhofer, R.~Weiss, V.~Dubourg, J.~Vanderplas, A.~Passos,
  D.~Cournapeau, M.~Brucher, M.~Perrot, and E.~Duchesnay, ``Scikit-learn:
  Machine learning in {P}ython,'' \emph{Journal of Machine Learning Research},
  vol.~12, pp. 2825--2830, 2011.

\bibitem{pereyra2017regularizing}
G.~Pereyra, G.~Tucker, J.~Chorowski, {\L}.~Kaiser, and G.~Hinton,
  ``Regularizing neural networks by penalizing confident output
  distributions,'' \emph{In International Conference on Learning
  Representations}, 2017.

\bibitem{1165342}
L.~{Rabiner} and B.~{Juang}, ``An introduction to hidden markov models,''
  \emph{IEEE ASSP Magazine}, vol.~3, no.~1, pp. 4--16, 1986.

\bibitem{JMLR:v11:raykar10a}
\BIBentryALTinterwordspacing
V.~C. Raykar, S.~Yu, L.~H. Zhao, G.~H. Valadez, C.~Florin, L.~Bogoni, and
  L.~Moy, ``Learning from crowds,'' \emph{Journal of Machine Learning
  Research}, vol.~11, no.~43, pp. 1297--1322, 2010. [Online]. Available:
  \url{http://jmlr.org/papers/v11/raykar10a.html}
\BIBentrySTDinterwordspacing

\bibitem{rodrigues2014gaussian}
F.~Rodrigues, F.~Pereira, and B.~Ribeiro, ``Gaussian process classification and
  active learning with multiple annotators,'' in \emph{International conference
  on machine learning}, 2014, pp. 433--441.

\bibitem{rodrigues2014sequence}
------, ``Sequence labeling with multiple annotators,'' \emph{Machine
  learning}, vol.~95, no.~2, pp. 165--181, 2014.

\bibitem{rodrigues2018deep}
F.~Rodrigues and F.~C. Pereira, ``Deep learning from crowds,'' in
  \emph{Thirty-Second AAAI Conference on Artificial Intelligence}, 2018.

\bibitem{sabetpour2020optsla}
N.~Sabetpour, A.~Kulkarni, and Q.~Li, ``Optsla: an optimization-based approach
  for sequential label aggregation,'' in \emph{Proceedings of the 2020
  Conference on Empirical Methods in Natural Language Processing: Findings},
  2020, pp. 1335--1340.

\bibitem{sang2003introduction}
E.~F. Sang and F.~De~Meulder, ``Introduction to the conll-2003 shared task:
  Language-independent named entity recognition,'' \emph{Proceedings of
  CoNLL-2003, Edmonton, Canada, 142 - 145}, 2003.

\bibitem{shelar2020named}
H.~Shelar, G.~Kaur, N.~Heda, and P.~Agrawal, ``Named entity recognition
  approaches and their comparison for custom ner model,'' \emph{Science \&
  Technology Libraries}, vol.~39, no.~3, pp. 324--337, 2020.

\bibitem{simpson-gurevych-2019-bayesian}
\BIBentryALTinterwordspacing
E.~D. Simpson and I.~Gurevych, ``A {B}ayesian approach for sequence tagging
  with crowds,'' in \emph{Proceedings of the 2019 Conference on Empirical
  Methods in Natural Language Processing and the 9th International Joint
  Conference on Natural Language Processing (EMNLP-IJCNLP)}.\hskip 1em plus
  0.5em minus 0.4em\relax Hong Kong, China: Association for Computational
  Linguistics, Nov. 2019, pp. 1093--1104. [Online]. Available:
  \url{https://www.aclweb.org/anthology/D19-1101}
\BIBentrySTDinterwordspacing

\bibitem{2008cheap}
R.~Snow, B.~O'Connor, D.~Jurafsky, and A.~Y. Ng, ``Cheap and fast --- but is it
  good? {E}valuating non-expert annotations for natural language tasks,'' in
  \emph{Proc.\ of the Conference on Empirical Methods in Natural Language
  Processing (EMNLP'08)}, 2008, pp. 254--263.

\bibitem{song2020learning}
H.~Song, M.~Kim, D.~Park, and J.-G. Lee, ``Learning from noisy labels with deep
  neural networks: A survey,'' \emph{arXiv preprint arXiv:2007.08199}, 2020.

\bibitem{sukhbaatar2014training}
S.~Sukhbaatar, J.~Bruna, M.~Paluri, L.~Bourdev, and R.~Fergus, ``Training
  convolutional networks with noisy labels,'' \emph{In International Conference
  on Learning Representations}, 2015.

\bibitem{tanno2019learning}
R.~Tanno, A.~Saeedi, S.~Sankaranarayanan, D.~C. Alexander, and N.~Silberman,
  ``Learning from noisy labels by regularized estimation of annotator
  confusion,'' in \emph{Proceedings of the IEEE Conference on Computer Vision
  and Pattern Recognition}, 2019, pp. 11\,244--11\,253.

\bibitem{tseng2001convergence}
P.~Tseng, ``Convergence of a block coordinate descent method for
  nondifferentiable minimization,'' \emph{Journal of optimization theory and
  applications}, vol. 109, no.~3, pp. 475--494, 2001.

\bibitem{wang2019symmetric}
Y.~Wang, X.~Ma, Z.~Chen, Y.~Luo, J.~Yi, and J.~Bailey, ``Symmetric cross
  entropy for robust learning with noisy labels,'' in \emph{Proceedings of the
  IEEE International Conference on Computer Vision}, 2019, pp. 322--330.

\bibitem{whitehill2009whose}
J.~Whitehill, T.-f. Wu, J.~Bergsma, J.~R. Movellan, and P.~L. Ruvolo, ``Whose
  vote should count more: Optimal integration of labels from labelers of
  unknown expertise,'' in \emph{Advances in neural information processing
  systems}, 2009, pp. 2035--2043.

\bibitem{yao2018deep}
J.~Yao, J.~Wang, I.~W. Tsang, Y.~Zhang, J.~Sun, C.~Zhang, and R.~Zhang, ``Deep
  learning from noisy image labels with quality embedding,'' \emph{IEEE
  Transactions on Image Processing}, vol.~28, no.~4, pp. 1909--1922, 2018.

\bibitem{yao2018online}
L.~Yao, L.~Su, Q.~Li, Y.~Li, F.~Ma, J.~Gao, and A.~Zhang, ``Online truth
  discovery on time series data,'' in \emph{Proceedings of the 2018 SIAM
  International Conference on Data Mining}.\hskip 1em plus 0.5em minus
  0.4em\relax SIAM, 2018, pp. 162--170.

\bibitem{YHY08}
X.~Yin, J.~Han, and P.~S. Yu, ``Truth discovery with multiple conflicting
  information providers on the web,'' \emph{IEEE Transactions on Knowledge and
  Data Engineering}, vol.~20, no.~6, pp. 796--808, 2008.

\bibitem{zhang2016learning}
J.~Zhang, X.~Wu, and V.~S. Sheng, ``Learning from crowdsourced labeled data: a
  survey,'' \emph{Artificial Intelligence Review}, vol.~46, no.~4, pp.
  543--576, 2016.

\bibitem{zhang2018generalized}
Z.~Zhang and M.~Sabuncu, ``Generalized cross entropy loss for training deep
  neural networks with noisy labels,'' in \emph{Advances in neural information
  processing systems}, 2018, pp. 8778--8788.

\bibitem{zheng2017truth}
Y.~Zheng, G.~Li, Y.~Li, C.~Shan, and R.~Cheng, ``Truth inference in
  crowdsourcing: Is the problem solved?'' \emph{Proceedings of the VLDB
  Endowment}, vol.~10, no.~5, pp. 541--552, 2017.

\bibitem{zhi2018dynamic}
S.~Zhi, F.~Yang, Z.~Zhu, Q.~Li, Z.~Wang, and J.~Han, ``Dynamic truth discovery
  on numerical data,'' in \emph{2018 IEEE International Conference on Data
  Mining (ICDM)}.\hskip 1em plus 0.5em minus 0.4em\relax IEEE, 2018, pp.
  817--826.

\bibitem{zhou2012learning}
D.~Zhou, S.~Basu, Y.~Mao, and J.~C. Platt, ``Learning from the wisdom of crowds
  by minimax entropy,'' in \emph{Advances in Neural Information Processing
  Systems (NIPS'12)}, 2012, pp. 2195--2203.

\end{thebibliography}


\end{document}